\documentclass[12pt]{article}

\makeatletter

\makeatletter
\def\@cite#1{$^{\mbox{\scriptsize{#1}}}$}
\def\@biblabel#1{(#1)}
\makeatother
 
\usepackage{graphicx}
\usepackage{float}
\restylefloat{figure}

\begin{document} 

\title{Hybrid Dynamic Density Functional Theory for Polymer Melts and Blends}

\author{
Takashi Honda \thanks{Author to whom correspondence should be addressed.} \\
Japan Chemical Innovation Institute, \\
and Department of Organic and Polymeric Materials, \\
Tokyo Institute of Technology,
Ookayama, Meguro-ku,
Tokyo 152-8552, Japan \\
\\
\and Toshihiro\ Kawakatsu
\\
Department of Physics, Tohoku University, \\
Aoba, Aramaki, Aoba-ku, Sendai 980-8578, Japan\\
}
\date{}
\maketitle




\begin{center}
{\Large\bf Abstract} \\
\end{center}

We propose a high-speed and accurate hybrid dynamic density functional theory for the computer simulations of the phase separation processes of polymer melts and blends.  The proposed theory is a combination of the dynamic self-consistent field (SCF) theory and a time-dependent Ginzburg-Landau type theory with the random phase approximation (GRPA).
The SCF theory is known to be accurate in evaluating the free energy of the polymer systems in both weak and strong segregation regions although it has a disadvantage of the requirement of a considerable amount of computational cost.  On the other hand, the GRPA theory has an advantage of much smaller amount of required computational cost than the SCF theory while its applicability is limited to the weak segregation region.
To make the accuracy of the SCF theory and the high-performance of the GRPA theory compatible, we adjust the chemical potential of the GRPA theory by using the SCF theory every constant time steps in the dynamic simulations. 
The performance of the GRPA and the hybrid theories is tested by using several systems composed of an A/B homopolymer, an AB diblock copolymer, or an ABC triblock copolymer.
Using the hybrid theory, we succeeded in reproducing the metastable complex phase-separated domain structures of an ABC triblock copolymer observed by experiments.
\newline

KEYWORDS: random phase approximation, self-consistent field theory, dynamic density functional theory, polymer phase separation

\newpage
\section{Introduction}

Polymer phase separation is an important problem in the practical applications of polymer materials composed of polymer alloys, polymer blends, and polymeric additives etc.
These materials show several types of phase separations, i.e., micro and macro phase separations, and phase separations induced by chemical reactions and external fields.
By controlling the morphology of the domains generated by the phase separations, we can design the physical properties of these polymeric materials such as permeability, electrical conductivity, and mechanical properties
\cite{Polymer Blends,Polymer Alloys and Blends}.
Therefore, understanding the mechanisms of the phase separation and the domain formation is very important in studying polymeric nanomaterials.  For this purpose, the density functional theories based on coarse-grained models of the polymer chains have been being used as important techniques
\cite{Fleer,Fredrickson,Matsen2003}.

In coarse-grained models of polymers, a polymer chain is modeled as a sequence of segments, which define the smallest length scale of the model
\cite{Kawakatsu book}.
In the density functional theories, the free energy of the system is expressed as a functional of the density distributions of the segments.
Using such an expression of the free energy, one can evaluate the chemical potential of each segment density.  In a diffusion dynamics, the segment density is driven by the spatial gradient of this chemical potential.
Depending on how the free energy is evaluated, the accuracy and the computational cost of the dynamic simulations are determined.  
The more the accuracy is improved, the more the computational cost is, in general, required. 
Therefore, it is important to establish a new technique where these two factors, i.e. the accuracy and the computational cost, are improved simultaneously.

A typical density functional theory that can give an accurate and reliable free energy is the self-consistent field (SCF) theory
\cite{Kawakatsu book,Helfand Wasserman,Hong}. 
In this SCF theory, the free energy and the chemical potential are accurately evaluated by taking account of the conformational entropy of the chains, which is described in terms of path integral technique under mean field approximation. 
A dynamic SCF theory is a dynamical extension of this SCF theory, where the diffusion dynamics of the segment densities is assumed
\cite{Kawakatsu book,Fraaije,Hasegawa Doi,Morita Kawakatsu,Zvelindovsky}. 
A simulation of this dynamic SCF theory requires recursive calculations to get the chemical potentials, which requires a considerable amount of computational cost.

There are other types of density functional theories which are based on simple phenomenological models of the free energy functional.  Typical examples are the time-dependent Ginzburg Landau theory and the Flory-Huggins-de Gennes theory
\cite{Kawakatsu book,Flory-Huggins-de Gennes,Oono Puri,Qi Wang,Kodama Doi,Nonomura Ohta,Yamada,Ren}.
In most of these theories, an analytical form of the free energy functional is assumed.  Using such a model free energy, one can evaluate the chemical potential without using the recursive calculations as is done in the SCF theory.  This accelerates the simulations considerably. 
In this article, we refer these theories as phenomenological density functional (PDF) theories.
In most of the PDF theories, the random phase approximation (RPA) plays a very important role
\cite{RPA de Gennes,de Gennes book}.
With this RPA, one can evaluate both the short range part and the long range part of the interaction energies between segment density fluctuations.

The Flory-Huggins-de Gennes' PDF theory uses a free energy model where the Flory-Huggins free energy for uniform systems is extended by introducing a square gradient term of the segment density fluctuations derived by the RPA
\cite{Flory-Huggins-de Gennes}.
This theory succeeded to reproduce the spatially inhomogeneous states of the phase separations of blends of linear homopolymers.
Leibler obtained a theoretical phase diagram of a diblock copolymer using a power series expansion of the free energy up to the fourth order terms in the segment density fluctuations, where the expansion coefficients are evaluated using the RPA
\cite{Leibler}.

Ohta and Kawasaki used a simpler form of the Leibler's free energy by decomposing it into a short-range part and a long-range part, and assumed a Ginzburg-Landau (GL) type form for the short-range part while the long-range part is approximated by the long-range asymptotic form of the second order term in the expansion
\cite{Ohta Kawasaki}. 
They succeeded in reproducing the phase diagram of the microdomain structures of diblock copolymer in the strong segregation region qualitatively.

Contrary to this Ohta-Kawasaki's PDF approach using an approximation on the second order term in the free energy expansion, Bohbot-Raviv and Wang evaluated the second order term faithfully using the RPA, and combined it with the higher-order terms in the expansion of the Flory-Huggins free energy. They predicted several microdomain structures of three-miktoarm star polymers and linear triblock copolymers using static calculations
\cite{Bohbot-Raviv Wang}.

Recently Uneyama and Doi proposed a PDF theory generalizing the Ohta-Kawasaki's theory and Flory-Huggins-de Gennes' theory.  Uneyama and Doi's model can be used for melts and blends of any types of polymer architectures, i.e. the branching structure and the order of the sequence of the segments along the chain
\cite{Uneyama Doi}.
They could reproduce micellar structures and obtained a phase behavior of the micelles that is in good agreement with the experimental phase diagram
\cite{Uneyama Doi 2}. 
They also applied their theory to dynamic problems associated with phase separations and structural phase transitions
\cite{Uneyama Ohta}.

Despite the success of these static and dynamic PDF theories, there are quantitative discrepancies between the results of the PDF theory and that of the SCF theory.  This is because of the truncation error of the Taylor series expansion of the free energy used in the PDF theories.  Such a truncation error cannot be negligible from the intermediate to the strong segregation regions.
On the other hand, the SCF theory is free from such a truncation error because the SCF theory relies on a numerical evaluation of the path integral, which corresponds to summing up all the terms in the Taylor series expansion of the free energy. Thus the SCF theory can give accurate results on the phase behavior, as long as the mean field approximation is valid, even in the strong segregation region where the truncation of the Taylor series expansion used in the PDF theories breaks down.

In references\cite{SUSHI Users Manual,G to C,book Honda Kawakatsu}, we applied the dynamic SCF theory to dynamics of phase separations and structural phase transitions, and evaluated the performance of the dynamic SCF theory.
In spite of the accuracy of the dynamic SCF theory, it requires a considerable amount of computational cost compared to the dynamic PDF theories. 
Thus a new technique or a new idea is needed to reduce the computational cost of the dynamic SCF theory.
For such a purpose, we first derive an exact expression for the scattering functions of the segment density fluctuations for polymer melts and blends composed of polymers with any architectures using RPA.
We do not rely on the approximate from of the free energy models proposed by Ohta and Kawasaki and by Uneyama and Doi
\cite{Ohta Kawasaki,Uneyama Doi}
but we use the free energy model proposed by Bohbot-Raviv and Wang
\cite{ Bohbot-Raviv Wang}
where the exact scattering function is used.
As the Bohbot-Raviv and Wang's model is based on the GL free energy using the RPA, we will abbreviate this model as GRPA.

Bohbot-Raviv and Wang applied this GRPA to static problems such as the equilibrium domain structures. 
In the present work, we apply this GRPA theory to the dynamic problems of polymer melts and blends.  We compare the results of the dynamic GRPA theory with those of the other PDF theories.

Based on the results of the dynamic GRPA theory, we next propose a hybrid theory where the dynamic SCF theory and the dynamic GRPA theory is combined to reduce the computational cost of the dynamic SCF theory without loss of its accuracy.
We test this hybrid theory by using several systems as examples.  We will show that the hybrid theory considerably accelerates the dynamic SCF theory without spoiling its accuracy.
Using this hybrid theory, we investigate the microphase-separation behavior of a linear ABC triblock polymer, which is known to show many complex domain structures by changing the affinity of the solvent to each of the blocks
\cite{Brinkmann,Abetz}.

For the present study, we used the "Simulation Utilities for Soft and Hard Interfaces (SUSHI)" in the OCTA system
\cite{SUSHI Users Manual}.

\section{Theory}

In this section, we first explain the detail of the static and dynamic GRPA theories.  The accuracy of the dynamic GRPA theory is then tested.  Based on these results, we propose a hybrid dynamic density functional theory that combines the dynamic SCF theory and the dynamic GRPA theory.

We consider a melt or a blend of polymers that have arbitrary molecular architectures.
(For simplicity, we assume that the polymers do not have loops.)
To construct a coarse-grained model, we adopt the Gaussian chain approximation where the polymers obey the Gaussian statistics on the scale larger than the segment size
\cite{Kawakatsu book}.

In our model, a polymer chain is divided into subchain(s), each of which is a Gaussian chain composed of the same type of segments.
For example, a homopolymer consists of a single subchain, and a block copolymer consists of subchains of different kinds. 
Here, the term "subchains" has the same meaning as a block in a block copolymer.
The molecular architecture of a polymer is defined in terms of its branching structures, the sequence of the subchains, the type of the segments that compose each subchain, and the length of each subchain.

We use a superscript index $p$ $(p = 1,2,3,\cdots, m)$ to identify the type of the polymer architecture where $m$ is the total number of polymer types, and a subscript index $K$ to identify the chemical type of the segments.
Subscript indices $i$ and $j$ $(i,j = 1,2,3,\cdots, n)$ are used to specify the subchains where we denote the number of subchains in a $p$-type polymer by $n^{(p)}$ and the total number of the subchain types in the system by $n=\sum_p n^{(p)}$.
We also define that the number of segments in the $i$-th subchain of a $p$-type polymer is $N_i^{(p)}$ and the total number of segments of the $p$-type polymer is $N^{(p)}=\sum_i N_i^{(p)}$.

Figure 1 shows a typical molecular architecture of a polymer chain (a $p$-type chain with $n^{p} = 5$).
Each subchain is specified by the number (i.e. the index $i = 1,2, \cdots, n^{p}$) and the chemical type of each subchain is distinguished using the line style.

We denote the interaction energy between a pair of $K$-type and $K'$-type segments by $\epsilon_{KK'}$.  This interaction energy is related to the Flory-Huggins interaction parameter $\chi_{KK'}$ by
\begin{equation}
\chi_{KK'} \equiv z \beta \{ \epsilon_{KK'} - \frac{1}{2} ( \epsilon_{KK} + \epsilon_{K'K'} ) \},
\end{equation}
where $z$ is the coordination number and $\beta = 1 / k_{\rm B} T $, $k_{\rm B}$ and $T$ being the Boltzmann constant and the temperature, respectively.

\subsection{Ginzburg-Landau free energy based on the random phase approximation}

\subsubsection{Free energy model of the system}
\label{Free energy model of the system}

During dynamic simulations, the system has not yet reached its equilibrium state.  This means that the segments in different subchains, in general, feel different chemical potentials even if the segments are of the same kind.
Thus, the free energy used in the dynamic simulations should be described in terms of the segment density fluctuations of each subchain defined by
\begin{equation}
\delta \phi_i({\bf r}) \equiv \phi_i({\bf r}) - \bar{ \phi_i },
\end{equation}
where $\phi_i({\bf r})$ is the local segment density of the $i$-th subchain at position ${\bf r}$, and $\bar{ \phi_i }$ is the average segment density of the $i$-th subchain.

Using the GL expansion that is justified in the weak segregation region, the free energy functional is given by
\cite{Kawakatsu book}
\begin{eqnarray}
{\cal F}[ \{ \phi_i ({\bf r}) \}] 
  &=& {\cal F}_0[ \{ \bar{ \phi_i } \} ] 
  + \frac{ 1 }{ 2 \beta } \sum_{ij}
      \int \int d{\bf r}\ d{\bf r}' S_{ij}^{-1} ( {\bf r}-{\bf r}' )
      \delta \phi_i({\bf r}) \delta \phi_{j} ( {\bf r}' )
  + ....
  \label{f1} \\
  &=& {\cal F}_0[ \{ \bar{ \phi_i } \} ] 
  + \frac{ 1 }{ 2 \beta } \sum_{ij}
      \int d{\bf q} S_{ij}^{-1} ( {\bf q} ) \delta \phi_i({\bf q}) \delta \phi_{j} ( - {\bf q} )
  + ....,
  \label{f2}
\end{eqnarray}
where the first term ${\cal F}_0[ \{ \bar{ \phi_i } \} ]$ is the free energy of the uniformly mixed state that is used as the reference state for the expansion.
Equations (\ref{f1}) and (\ref{f2}) are the Taylor series expansions of the free energy with respect to the local segment density fluctuations of the subchains. 
The expansion coefficients, $S_{ij}^{-1} ( {\bf r}- {\bf r}' ), $ in eq (\ref{f1}) is the inverse of the density-density auto-correlation function between the segment density fluctuations belonging to the $i$-th and $j$-th subchains at positions ${\bf r}$ and ${\bf r}'$, respectively.  Its Fourier representation is $S_{ij}^{-1} ( {\bf q} )$ in eq~(\ref{f2}) where ${\bf q}$ is the scattering wave vector.
The dots at the ends of eqs~(\ref{f1}) and (\ref{f2}) mean the higher order terms in the Taylor series expansion in $\delta \phi_i({\bf r})$ or $\delta \phi_i({\bf q})$, respectively.

Using this GL expansion, Bohbot-Raviv and Wang tried to include the effects of the spatial inhomogeneity into the Flory-Huggins free energy for uniform systems.  They replaced the second order contributions in $\delta\phi_i({\bf r})$ in the Flory-Huggins free energy with the second term of eq (\ref{f2}), and they obtained the following expression
\cite{Bohbot-Raviv Wang}.
\begin{eqnarray}
{\cal F}[ \{ \phi_i ({\bf r}) \}] 
&=& \frac{1}{\beta} \sum_i \int d{\bf r} \frac{ \phi_i({\bf r}) }{ N_i} \ln \phi_i({\bf r}) 
      - \frac{1}{2 \beta} \sum_i \int d{\bf r} \frac{1}{ N_i  \bar{ \phi_i } } \delta \phi_i({\bf r}) \delta \phi_i({\bf r}) 
\nonumber \\
& & + \frac{1}{2\beta} \sum_{ij} \int d{\bf q} S_{ij}^{-1} ( {\bf q} ) \delta \phi_i({\bf q}) \delta \phi_{j} ( - {\bf q} ),
\label{f3}
\end{eqnarray}
where $N_i$ is the length of the $i$-th subchains (the superscript $p$ is suppressed for simplicity).
In eq (\ref{f3}), the first term is the Flory-Huggins mixing entropy of the centers of mass of the subchains.  When this mixing entropy is expanded in a power series in $\delta\phi_i({\bf r})$, the second order term is given by the second term of eq (\ref{f3}).  By subtracting this contribution and instead adding the more accurate expression, i.e. the third term that contains $S_{ij}^{-1} ( {\bf q} )$, one can generalize the Flory-Huggins free energy to inhomogeneous systems.  In this model, it is essentially important in simulating the strong segregated systems that we retain the higher order contributions in the expansion of the free energy.  
In eq (\ref{f3}), we can neglect the constant term and the terms linear in $\delta\phi_i({\bf r})$ because these terms do not affect the phase separation dynamics.  Thus, the second order terms are the leading non-trivial contributions in the GL expansion.

Uneyama and Doi have generalized the third term of eq (\ref{f2}) by using the approximate equations of the scattering functions derived by Ohta and Kawasaki 
\cite{Uneyama Doi} as
$
S_{ij}^{-1}({\bf q}) = A/{\bf q}^2 + B + C {\bf q}^2,
$
where $A$, $B$, and $C$ are fitting parameters.
On the other hand, we use the exact expression of the $S_{ij}^{-1}({\bf q})$ obtained using the RPA
\cite{Leibler}.

To obtain the exact expression of $S_{ij}^{-1}({\bf q})$, the linear relation $u_i({\bf q}) = \sum_j S_{ij}^{-1}({\bf q})\delta\phi_j({\bf q})$ is assumed in the RPA, where $S_{ij}^{-1}({\bf q})$ is the $ij$-element of the inverse matrix of the scattering function matrix of segment density fluctuations.
Due to the incompressibility condition, one of the elements of $\{ \delta\phi_i \}$ is not independent of the other.
This leads to the use of a reduced scattering function matrix $\{ \widetilde{S}_{ij}^{-1}({\bf q}) \}$ instead of $\{ S_{ij}^{-1}({\bf q}) \}$ itself.  The detail of the derivation of this scattering function matrix is given in Appendix \ref{Generalized scattering function}.

With this $\{ \widetilde{S}_{ij}^{-1}({\bf q} \})$, we can explicitly calculate the external potential field $\{ u_i({\bf r}) \}$ without relying on any recursive calculations when the segment density distribution $\{ \delta\phi_i({\bf r}) \}$ is given.

\subsubsection{Free energy and chemical potential}

By using $\{ \widetilde{S}_{ij}^{-1}({\bf q}) \}$, the free energy of the system eq (\ref{f3}) is rewritten as
\begin{eqnarray}
{\cal F}[ \{ \phi_i ({\bf r}) \}] 
&=& \frac{1}{\beta} \sum_i^n \int d{\bf r} \frac{ \phi_i({\bf r}) }{ N_i} \ln \phi_i({\bf r}) 
  - \frac{1}{2 \beta} \sum_i^n \int d{\bf r} \frac{1}{ N_i  \bar{ \phi_i } } \{ \delta \phi_i({\bf r}) \}^2 
\nonumber \\
& & + \frac{1}{2} \sum_{i}^{n} \int d{\bf r} u_i({\bf r}) \delta \phi_i({\bf r}).
\label{free energy of the system}
\end{eqnarray}
The coefficients $\{ u_i({\bf r}) \}$ in the third term of eq~(\ref{free energy of the system}) are calculated by using the matrix $\{ \widetilde{S}_{ij}^{-1} \}$.

Using the free energy model eq \ref{free energy of the system},
the chemical potential of the segments of the $i$-th subchain is given by
\begin{eqnarray}
\tilde{\mu}_i({\bf r}) 
&=& \frac{ \delta {\cal F}[ \{ \phi_i ({\bf r}) \}]  }{\delta \phi_i({\bf r}) } \\
&=& \frac{1}{\beta} \Bigl \{
         \frac{ \ln \phi_i({\bf r}) }{ N_i } + \frac{ 1 }{ N_i } 
       - \frac{ 1 }{ \bar{ \phi_i } N_i } \delta \phi_i({\bf r})
\nonumber \\
& &    -\frac{ \ln \phi_n({\bf r}) }{ N_n } - \frac{ 1 }{ N_n } 
       + \frac{ 1 }{ \bar{ \phi_n } N_n } \delta \phi_n({\bf r})
                    \Bigr \}
       + u_i({\bf r}),
\label{chemical potential}
\end{eqnarray}
where the incompressibility condition is used. Under such an incompressibility condition, the $n$-th elements $\phi_n({\bf r})$ and $\delta \phi_n({\bf r})$ are treated as dependent variables as
\begin{eqnarray}
       \phi_n({\bf r}) &=& 1 - \sum_i^{n-1} \phi_i({\bf r})\\
\delta \phi_n({\bf r}) &=& - \sum_i^{n-1} \delta \phi_i({\bf r}).
\label{incompressibility condition of phi}
\end{eqnarray}
These $n$-th elements and $\tilde{\mu}_n({\bf r})$ are not explicitly used in the dynamic calculations.

\subsection{Dynamic density functional theory}

The time evolution equation for the segment density of the $i$-th subchain can be described in terms of the chemical potential $\tilde{\mu}_{}({\bf r})$ under the assumption of Fick's law of linear diffusion for the segment densities as follows
\begin{equation}
  \frac{\partial}{\partial t} \phi_i({\bf r},t) 
       =  \nabla \cdot L_i({\bf r},t) \nabla \tilde{\mu}_{}({\bf r},t),
\label{dynamic equation of diffusion}
\end{equation}
where $t$ is the time and $L_i({\bf r},t)$ is the mobility of the segment of the $i$-th subchain at position ${\bf r}$ at time $t$.
We use the following density dependent mobility
\begin{equation}
  L_i({\bf r},t) = L_i^0 \phi_i({\bf r},t),
\label{Mobility}
\end{equation}
where $L_i^0$ is a constant value.
This assumption prevents the segment density $\phi_i({\bf r},t)$ from being negative value.  It is useful because the Taylor series expansion in terms of the density fluctuation $\delta \phi_i({\bf r},t)$ around $\bar{\phi}_i$ does not guarantee the positiveness of the values of $\phi_i({\bf r},t)$.

A variable time mesh technique, whose detail is described in Appendix \ref{Variable time mesh technique for the dynamic density functional theory}, is also adopted to avoid the negative values of the segment density $\phi_i({\bf r},t)$.

\subsection{Accuracy of the dynamic GRPA theory}

We compare the domain structures of several block copolymer systems obtained using the dynamic GRPA theory with those obtained using some other theories.

\subsubsection{One dimensional lamellar structure of a symmetric diblock copolymer}

Figure 2 shows the lamellar structure of a symmetric AB diblock copolymer (block ratio $f=0.5$ ) melt obtained with the dynamic GRPA theory. The size of the simulation box is optimized using the dynamic system size optimization (SSO) method
\cite{G to C},
where the GRPA free energy is used in determining the rate of change of each side length of the simulation box.

Figure 2a shows the equilibrium period of the lamellar structure as a function of $\chi N$ where $N$ is the total chain length. 
The result of the GRPA theory lies between that of the SCF theory and that of Uneyama and Doi's theory. Uneyama and Doi's theory is similar to our theory but uses the approximate form of the scattering function, which causes the disagreement. As the GRPA free energy coincides with the SCF free energy up to the second order in its Taylor series expansion, the behavior of these two models should be the same when the system approaches the critical point.  We can actually confirm this in Fig.2a, i.e., the results of the GRPA and the SCF theories converge to the same result near the theoretical critical point ($\chi N = 10.5$ for symmetric block copolymer).

Figure 2b shows the profiles of an interface of an A/B polymer blend for different values of $\chi N$ obtained by using GRPA and SCF theories.  The interfacial profiles obtained with the GRPA theory are always sharper than those obtained with the SCF theory.
The origin of such a sharp interfacial profile is expected to be the approximate form of the higher order terms in the expansion of the GRPA free energy where the effects of the chain conformation are neglected.
The available region of the GRPA calculation extends up to about $\chi N=50$ at which point the interfacial profile becomes too sharp to be simulated with a reasonable mesh size compared to that of the dynamic SCF theory.

\subsubsection{Three dimensional cylindrical structure of a diblock copolymer}
\label{Three dimensional cylinder structure of a diblock copolymer}

We tested the hybrid theory using a block copolymer system with a set of parameters which corresponds to the hexagonally packed cylinder (HEX) as the equilibrium state
\cite{book Honda Kawakatsu}.
Figure 3 shows the comparison of the obtained structures between the dynamic GRPA theory (Fig. 3a) and the dynamic SCF (Fig. 3b) theory. 
The dynamic GRPA theory can not reach the equilibrium HEX structure, while the dynamic SCF theory can.
Such a discrepancy is caused by the difference of the free energy model between these two theories. 
The structure shown in Fig. 3a has many three fold junctions, which are characteristic to the bicontinuous double gyroid (G) structure. We compared the densities of the GRPA free energy for the HEX structure (Fig. 3b) and for the G structure
\cite{G to C}
at this parameter set ($f=0.35$ and $\chi N = 15$).  This comparison showed that the G structure is more stable than the HEX structure for the GRPA model.  This is due to the fact that the phase diagrams of diblock copolymers are quantitatively different between the SCF and GRPA theories.
Thus the GRPA theory can not be used for quantitative evaluations of the phase diagram while the SCF theory can.

\subsection{Hybrid dynamic density functional theory}

The discrepancy between the GRPA and SCF theories as described in the preceding section is caused by the assumption in the evaluation of the higher order terms in the Taylor series expansion in the GRPA theory. This is because the accuracy of the evaluated higher order terms is depended on the treatment of the conformational entropy of the chains.  The GRPA theory neglects the conformation entropy in the higher order terms while the SCF theory evaluates them correctly.

To improve the accuracy of the dynamic GRPA theory up to the level of the dynamic SCF theory, one can combine these two theories.
This can be achieved by adding a correction term $\Delta \bar{\mu}_ ({\bf r})$ to the chemical potential of the GRPA theory $\tilde{\mu}_i^{RPA}({\bf r},t)$ given in eq (\ref{chemical potential}) as
\begin{equation}
  \tilde{\mu}_i^{HYB}({\bf r},t) = \tilde{\mu}_i^{RPA}({\bf r},t) + \Delta \bar{\mu}_i ({\bf r},t),
\end{equation}
where the $\tilde{\mu}_i^{HYB}({\bf r},t)$ means a corrected chemical potential. 
The correction term $\Delta \bar{\mu}_i ({\bf r})$ is defined as the difference between the chemical potential obtained with the SCF theory $\tilde{\mu}_i^{SCF}({\bf r},t)$  and that obtained with the GRPA theory $\tilde{\mu}_i^{RPA}({\bf r},t)$. 
Here we evaluate $\Delta \bar{\mu}_i ({\bf r})$ at every constant intervals of time $t_H$ as
\begin{eqnarray}
  \Delta \bar{\mu}_i ({\bf r},t) &=& \tilde{\mu}_i^{SCF}({\bf r},n t_H) - \tilde{\mu}_i^{RPA}({\bf r}, n t_H)\ \ \ \ \ {\rm for\ }\ \ nt_H \le t \le (n+1)t_H.
\label{delta Mu 2}
\end{eqnarray}
where $n$ is a positive integer.
This correction procedure hybridizes the dynamic SCF and dynamic GRPA theories.

Using this corrected chemical potential $\tilde{\mu}_i^{HYB}({\bf r},t)$, we can perform dynamic simulations. 
Although the speed of the calculation of the hybrid theory is slower than that of the dynamic GRPA theory, it is much faster than that of the dynamic SCF theory.
The hybrid theory guarantees that the final equilibrium structure (or the steady state in the late stage) is the same as that produced in the late stage of the corresponding dynamic SCF simulation.
If we choose the interval of the update of $\Delta \bar{\mu}_i ({\bf r})$ short enough, the result of the hybrid theory should trace that of the dynamic SCF theory.  Therefore, the interval of update $t_H$ is a crucial parameter for the accuracy of the hybrid theory.
In practical calculations, the simulation time step is discretized with $\Delta t$.
Thus the number of time steps between the consecutive updates of $\Delta \bar{\mu}_i ({\bf r})$ is denoted by $n_H = t_H / \Delta t$.
Hereafter, we will use this $n_H$ to indicate the degree of the hybridization.

\section{Simulation Results}

In this section, we will test the accuracy and the efficiency of the hybrid theory by comparing its results with those of the dynamic GRPA and dynamic SCF theories using several examples.  

In Sec.~\ref{A/B polymer blend}, we check the quantitative accuracy of the hybrid theory by comparing the time dependences of the scattering function for various values of $n_H$ using a two dimensional symmetric A/B polymer blend that undergoes spinodal decomposition as a target system.  We will see that the hybrid theory with a rather large value of $n_H$ can reproduce the results of the correct SCF simulation well.

If one wants to obtain a qualitative (not quantitative) phase diagram of a given polymer system, a correct evaluation of the relative order of the values of the free energy for different microdomain structures is important.  In Sec.~\ref{Hexagonally packed cylinder structure}, we evaluate the free energy of a three dimensional cylindrical structure of an asymmetric A-B diblock copolymer using both the GRPA and the SCF theories, and compare their results to show that these two theories give qualitatively the same results.

The accuracy and the efficiency of the hybrid theory is expected to be strongly dependent of the choice of the correction interval $n_H$.  We check such an effect in Sec.~\ref{Computational efficiency of the hybrid theory}.  We will confirm that the hybrid theory is efficient and correct even with a large value of $n_H$.
However, the hybrid theory is not always efficient.  Such an exceptional example where the hybrid theory cannot trace the correct dynamics will be shown in Sec.~\ref{Exception of the hybrid theory}.

To show the usefulness of the hybrid theory, in Sec.~\ref{Morphology of a ABC triblock copolymer controlled by solvent selectivity},
we show a results of the hybrid simulation on a formation of complex microphase-separated domains of a cast film of an ABC triblock copolymer where the micro domain structures are controlled dynamically through the evaporation of a selective solvent
\cite{Brinkmann,Abetz}.

In all the simulations shown in this section, the effective bond lengths of all segments are assumed to be the same value $b$, which we use as the unit of length.

\subsection{Quantitative accuracy of the hybrid theory: A/B polymer blend case}
\label{A/B polymer blend}

Here, we test the quantitative accuracy of the hybrid theory by using a two dimensional A/B polymer blend system which undergoes a spinodal decomposition as an example.
Figure 4 shows the time dependence of the wave number of the peak position of the circularly averaged scattering function.
The wave number at the peak is obtained by fitting the scattering function by a Gaussian function using the least square method with 20 data points near the peak position (the range of ${\bf q}$ is 0.078 $b^{-1}$).
The values of $n_H$ are shown in the figure.

The time dependence of the peak position of the scattering function for the four cases with SCF( $n_{H}=$ 1 ), hybrid( $n_{H}=$ 200, 1600 ) and GRPA( $n_{H}= \infty$) are plotted in Fig. 4.
Here, the two curves for SCF and $n_{H}=200$ are almost overlapping and the curve for $n_{H}=1600$ is also very close to these curves.
There is a tendency that the wave number of the dynamic GRPA theory is larger than that of the dynamic SCF theory, i.e., the GRPA theory underestimates the domain size than the SCF theory.
This feature is the same as the one dimensional calculation result of lamellar as shown in Fig. 2a.
On the other hand, it is surprising that the wave number of the hybrid theory with even a large value of $n_H$, i.e. $n_{H}=1600$ coincides well with that of the dynamic SCF theory.
This result demonstrates that the hybrid theory can trace the same morphological change of domains as the dynamic SCF theory.
A further increase in $n_{H}$ such as 3600 prevents the SCF iteration scheme from converging because the interfaces become too sharp in the late stage of the dynamic GRPA simulation.

\subsection{Qualitative accuracy of the phase behavior: Hexagonally packed cylinder case}
\label{Hexagonally packed cylinder structure}

For a qualitative understanding of the phase behavior of a given system, it is essential that the theory can give correct order of the values of the free energy for different microdomain structures.  Here, we check whether the GRPA theory can give a qualitatively the same result as the SCF theory which is regarded as a correct reference value.
For this purpose, we perform hybrid simulations with different $n_H$ values on the formation process of the HEX structure shown in Fig. 3b starting from an initial disordered state.
For each simulation run, we evaluate the free energy based on the calculated segment density fields.

Figure 5 shows the time dependence of the free energy density for various values of $n_H$.
The upper set of curves shows the free energy densities evaluated with the SCF theory while the lower set of curves shows those evaluated with the GRPA theory.   In both sets, the segment density distributions are calculated with the hybrid theory with the $n_H$ vales shown in the figure.
In all cases, the free energy converges to the same value in the late stage when the system reaches the equilibrium HEX phase as shown in Fig. 3b. 
Although the absolute values of the free energy are different for the two cases, i.e. SCF and GRPA theories, the overall and relative behaviors of the individual runs are almost the same for these tow sets of curves.
This result means that the computationally economical GRPA theory can be used in evaluating the free energy instead of the computationally expensive SCF theory as long as the dynamics is traced using the hybrid simulation method.  Thus, a combination of the hybrid dynamical simulations and the free energy evaluation using GRPA theory is a useful technique to predict the (qualitative) non-equilibrium phase diagram. (Or we should call it as "stability diagram " because the system is not in the equilibrium state even in the final stage.)

\subsection{Computational efficiency of the hybrid theory}
\label{Computational efficiency of the hybrid theory}

Figure 6 shows the relation between the $n_H$ and the computational time required for the hybrid simulation. Here, the computational time is normalized with that of the dynamic SCF theory.
Results for three different systems are shown; a one dimensional symmetric A-B diblock copolymer melt in the lamellar phase, a two dimensional symmetric A/B polymer blend which undergoes a spinodal decomposition (the same system as was shown in Fig. 4), and a three dimensional asymmetric A-B diblock copolymer melt in the HEX phase (the same as that shown in Figs. 3 and 5).

The required computational time decreases with increasing $n_H$. This is not trivial because increasing $n_H$ in general results in a larger difference between the GRPA result and the correct SCF result after $n_H$ time steps, which leads to an increased number of iterations in the SCF calculation. 
The reduction rate depends on the simulation conditions, i.e. the system size, the number of components, and the molecular architecture of the polymers. 
In the same figure, the computational times required for the dynamic GRPA simulations ($n_H=\infty$) are also shown with the horizontal lines, which give the lower limits for the computational times for the hybrid theory.
These results show that the computational efficiency of the hybrid theory quickly approaches its optimum value (the GRPA case) when the value of $n_H$ is larger than a critical value.

\subsection{Inefficient case for the hybrid theory}
\label{Exception of the hybrid theory}

There are certain systems where a very small value of $n_H$ ($n_H \sim 5$) is required for the hybrid theory to trace the result of the dynamic SCF theory.
Such an inefficiency is caused by the discrepancy between the equilibrium segment density profiles calculated with the GRPA and the SCF theories.

As an example, Fig. 7 shows the one-dimensional segment density profiles of a blend of an ABA triblock copolymer and a B homopolymer.  Figure 7a is the profile obtained with the dynamic GRPA theory while Fig. 7b is that obtained with the SCF theory.
In Fig. 7a, an accumulation of the A segments and the B segments of the ABA copolymer is observed at the interfaces between the domains of the B segments of the B homopolymer. There is no internal fine structure in such interfacial regions.  On the other hand, in Fig.~7b, one can observe a microphase separation of the A segments and the B segments of the ABA copolymer inside these interfacial regions.  
Such a discrepancy means that the GRPA theory cannot reproduce the fine structure in the interfacial region.  Thus, in order for the hybrid theory to trace the correct SCF results, one has to use a small $n_H$ to keep track of the change in the correct segment profile of the SCF theory at any time step of the dynamical simulations.

Even with a small value of $n_H$, the efficiency of the hybrid theory is still better than that of the full SCF theory ($n_H = 1$) as was discussed in Fig.~6. 
Figure 8 shows a result of a hybrid simulation with $n_H = 5$ on the same system as was shown in Fig. 7.
The ABA copolymer forms micelles in the matrix of the B homopolymer.  The micelles have the core-shell structure of which the core and shell are formed by the A and B segments of the ABA polymer, respectively.
Such a complex domain structure can be obtained with the hybrid theory with a considerably improved efficiency compared to the SCF theory.

\subsection{Morphology of an ABC triblock copolymer controlled by solvent selectivity}
\label{Morphology of a ABC triblock copolymer controlled by solvent selectivity}

It is known that a cast film of polystyrene-$block$-polybutadiene-$block$-poly(methyl methacrylate) (SBM) solvated with a tetrahydrofuran (THF) solution often shows a structural motif, i.e. a characteristic microdomain formation.
For example, in refs.~\cite{Brinkmann,Abetz}, a HEX structure of the triblock copolymer is controlled by changing the selectivity of the THF solvent to each of the blocks.
In the process of the evaporation of the THF, HEX domains composed of the poly(methyl methacrylate) (PM) first appear, where the PS and PB blocks make a uniform matrix phase.  After such HEX domains are formed, the PS segments start to accumulate around the HEX domains of PM and form a structural motif.

We simulated this phenomenon using a hybrid simulation on a two dimensional ABC triblock copolymer system.
We assume that the SBM triblock cpolymer is modeled with an $A_{20}B_{60}C_{20}$ triblock cpolymer.  The evaporation of the THF solvent is effectively modeled as a change in the value of the $\chi_{ij}$ parameters.
The simulation results are shown in Fig.~9, where the segment density of the A segments are shown.  Figure 9(a) shows the initial HEX structure of the C-segments, where the $\chi$ parameters are chosen as $\chi_{AB} = 0.0$ and $\chi_{AC} = \chi_{BC} = 0.4$, respectively.
In this initial state, the A and B segments are dissolved in the THF solvent to form a matrix.  Starting from this initial HEX structure, a domain formation of the A (PS) segments is simulated by changing the $\chi$-parameters to $\chi_{AB} = \chi_{AC} = \chi_{BC} =0.4$, which corresponds to a change in the quality of the THF solvent by a solvent evaporation.
Figures 9b and 9c show the time dependent distributions of the A segments.
The A segments migrate in the matrix as Fig 9b and make a hexagonally arranged structural motif as shown in Fig. 9c.  To show the fine structure of the system in Fig.~9c, we show in Fig.~9d the same system as in Fig. 9c but shows the final segment density profiles of both the A and the C segments.  We can confirm that the system is composed of a HEX domains of the C segments surrounded by the motif of the A segments
\cite{Brinkmann,Abetz}.

A static calculation starting from the disordered state can not in general produce the final structure shown in Fig. 9d because the equilibrium state of the symmetric $A_{20}B_{60}C_{20}$ triblock copolymer with $N_A=N_C$ and $\chi_{ij}=0.4$ should be a symmetric lamellar structure (ABCBA\ldots) with respect to the exchange between the A and the C segments. 
Therefore the structure of Fig. 9d must be a metastable structure generated in the dynamic process with the THF evaporation.
This simulation demonstrates the usefulness of the hybrid theory, which can reproduce a complex domain formation process with a considerably improved calculation efficiency than the dynamic SCF theory.

\section{Discussion}

The GRPA theory is based on both the Flory-Huggins theory and the RPA up to the second order terms in the series expansion of the free energy.
These assumptions cause a deviation of the chemical potential from that of the SCF theory.
Although the dynamic simulation based on the GRPA theory can follow the dynamic SCF calculation qualitatively as shown in Fig. 4, the result of the GRPA theory is quantitatively inadequate.
The hybrid theory can correct this deviation and reproduces the result of the SCF theory as shown in Figs. 4 and 5.

Our GRPA theory uses the same free energy model of Bohbot-Raviv and Wang's theory
\cite{Bohbot-Raviv Wang}, where we use the exact expressions of the scattering functions derived using the RPA while Ohta-Kawasaki
\cite{Ohta Kawasaki}
and Uneyama and Doi's
\cite{Uneyama Doi}
theories use approximate expressions of the scattering functions.
As Bohbot-Raviv and Wang reported the results only for static and two dimensional systems, we performed three dimensional dynamic calculation for polymer melts and blends of any polymer architectures using our GRPA theory.

Uneyama and Doi have reported that the Bohbot-Raviv and Wang's theory cannot give the correct dependence of the interfacial tension of a block copolymer melt on its block ratio
\cite{Uneyama Doi}.
Moreover, both of the dynamic GRPA theory and the Uneyama and Doi's theory show a considerable deviation from the result of the SCF theory as shown in Fig. 2a, especially in the strong segregation region.
These difficulties are eliminated with the use of the hybrid theory.
The hybrid theory also improves the computational efficiency which is discussed in Appendix \ref{Efficiency of the hybrid theory}.

\section{Conclusion}

We proposed a hybrid theory for the dynamics of the phase separation phenomena of polymer melts and blend systems composed of polymers with any molecular architectures.
To improve the efficiency and accuracy, the dynamic SCF and dynamic GRPA theories are hybridized.  The former is quantitatively accurate while the latter has an advantage in the computational efficiency.
Although the parameters required for the GRPA theory are the same as those for the SCF theory, the GRPA theory demands neither the recursive calculations nor any extra model parameters such as the empirical parameters required for the GL free energy.
In the hybrid theory, the time evolution of the segment density fields is mainly calculated using the chemical potential of the GRPA theory, which is corrected using the chemical potential of the SCF at some intervals.

We confirmed that this hybrid theory accelerates the dynamic SCF theory without loss of its accuracy by using several polymer melts and blend systems.  We also confirmed that the dynamic hybrid theory well reproduces the morphological changes of domains quantitatively.  Thus, this hybrid theory will enable us to perform accurate large scale simulations on polymer melts and blend systems.

The GRPA theory requires proper boundary conditions which is adapted to the Fourier transformation because the scattering functions, which are the coefficients of the series expansion of the free energy, are calculated in the Fourier space. This condition makes it difficult to apply the dynamic GRPA and hybrid theories to the systems without periodic boundary conditions, such as a system subjected to a solid wall.
\\
\\
{\Large\bf Acknowledgment}
\\

The authors thank Prof. M. Doi (Tokyo University) and T. Uneyama (Kyoto University) for many comments and discussions.
This study is executed under the national project on nanostructured polymeric materials, which has been entrusted to the Japan Chemical Innovation Institute by the New Energy and Industrial Technology Development Organization (NEDO) under METi's Program for the Scientific Technology Development for Industries that Create New Industries.
This work is partially supported by Grant-in-Aid for Science from the Ministry of Education, Culture, Sports, Science and Technology, Japan. 
\newline

\appendix

{\Large\bf Appendices}

\section{Scattering functions based on RPA for polymers with any molecular architectures}
\label{Generalized scattering function}

In this appendix, we derive an exact expression of the inverse of the scattering function matrix ${\bf S}^{-1}({\bf q}) \equiv \{ S_{ij}^{-1}({\bf q})\}$.  
For the simplicity of the descriptions, we define vectors of the Fourier modes of the segment density fluctuations of each subcahin and the external potentials acting on each segment density as:
\begin{eqnarray}
{\bf x}({\bf q}) \equiv \{ \delta \phi_i( {\bf q} ) \} \\
{\bf u}({\bf q}) \equiv \{ u_i( {\bf q} ) \}.
\end{eqnarray}
We also define the same vectors but in the real space as:
\begin{eqnarray}
{\bf x}({\bf r}) \equiv \{ \delta \phi_i( {\bf r} ) \} \\
{\bf u}({\bf r}) \equiv \{ u_i( {\bf r} ) \},
\end{eqnarray}
where ${\bf r}$ is the position vector.
A linear relation ${\bf x}({\bf q}) = {\bf S}({\bf q}){\bf u}({\bf q})$ between these two vectors is assumed, where the coefficient matrix ${\bf S}({\bf q})$ corresponds to the inverse of ${\bf S}^{-1}({\bf q})$. 
Due to the incompressibility condition, one of the elements of ${\bf x}$ is not independent to the other, and therefore ${\bf S}({\bf q})$ is a singular matrix.
Thhesn, in order to obtain ${\bf S}^{-1}({\bf q})$, we introduce $\widetilde{{\bf S}}({\bf q})$ where one of the rows and corresponding column of ${\bf S}({\bf q})$ are eliminated so that $\widetilde{{\bf S}}^{-1}({\bf q})$ can exist.

We introduce a matrix whose elements are the scattering functions of ideal Gaussian chains as 
\begin{equation}
{\bf S }^0 \equiv \{ S_{ij}^0( {\bf q} ) \},
\end{equation}
where $S_{ij}^0( {\bf q} )=0$ if the $i$-th and the $j$-th subchains that belong to different polymers. The explicit expressions of the elements $ S_{ij}^0( {\bf q} )$ are given later.
We also define a matrix of the segment interaction energy as
\begin{equation}
{\bf C } \equiv \{ z \epsilon_{ij} \},
\end{equation}
where $\epsilon_{ij}$ means the interaction energy between segments of the $i$-th and $j$-th subchains.
Hereafter, the expressions ${\bf x }$ and ${\bf u }$ without arguments are only used for those in the Fourier space.

The incompressibility condition generates a static pressure that is common to all the subchains.  We denote this static pressure in the Fourier space as $u^*({\bf q })$.  This $u^*$ is a function of ${\bf u },{\bf S }^0$ and ${\bf C }$ as
\begin{equation}
u^* = f({\bf u }, {\bf S }^0, {\bf C }).
\end{equation}
Using the quantities introduced above, the self-consistent equation is given by
\begin{equation}
{\bf x } = - \beta {\bf S }^0({\bf u } + {\bf C }{\bf x } + u^*{\bf e} ),
\label{generalized equation of RPA}
\end{equation}
where ${\bf e}$ is the vector whose all elements are unity.
This equation is the fundamental equation of the generalized RPA that gives the coefficient of the second order term in the free energy expansion.
According to Leibler, the sum of terms in parentheses on the right-hand side of eq (\ref{generalized equation of RPA}) is called the effective potential that is an external potential renormalized by the internal interactions and the incompressibility condition
\cite{Leibler}.
Solving eq (\ref{generalized equation of RPA}) with respect to ${\bf x }$ gives
\begin{eqnarray}
{\bf x } 
&=&
- \beta ( {\bf E } + \beta {\bf S }^0{\bf C })^{-1}
  {\bf S }^0({\bf u }+u^*{\bf e}) \nonumber \\
&=&
- \beta ({\bf Bu }+u^*{\bf Be}),
\label{equation of p}
\end{eqnarray}
where ${\bf E }$ is the identity matrix and ${\bf B }=({\bf E }+ \beta {\bf S }^0{\bf C })^{-1}{\bf S }^0$.
Here the incompressibility condition must be satisfied for ${\bf x }$ as 
\begin{equation}
\sum_{i=1}^n x_i = 0.
\label{incompressibility condition of x}
\end{equation}
From eqs (\ref{equation of p}) and (\ref{incompressibility condition of x}), $u^*$ is obtained as
\begin{eqnarray}
u^*
 &=& -\ \frac{\displaystyle \sum_{i=1}^n \sum_{j=1}^n B_{ij}u_j }
            {\displaystyle \sum_{i=1}^n \sum_{j=1}^n B_{ij}    }.
\label{v_RPA}
\end{eqnarray}
Substituting eq (\ref{v_RPA}) into eq (\ref{equation of p}) leads to
\begin{eqnarray}
{\bf x }  
          =  - \beta {\bf S } {\bf u },
\label{exact RPA}
\end{eqnarray}
where ${\bf S } = {\bf B } + {\bf B }'$ and the element of the matrix ${\bf B }'$ is defined as
\begin{equation}
B'_{ij} = -\ \frac{ \displaystyle \Biggl( \sum_{j'=1}^n B_{ij'} \Biggr) 
                                 \Biggl( \sum_{i'=1}^n B_{i'j} \Biggr) }
                 { \displaystyle \sum_{i'=1}^n \sum_{j'=1}^n B_{i'j'} }.
\end{equation}

If the inverse of ${\bf S }$ can be obtained in eq~(\ref{exact RPA}), we can get ${\bf u } = (-1/\beta){\bf S }^{-1}{\bf x }$ and this ${\bf u }$ can be used for the calculation of the free energy.
However, due to the incompressibility condition, eq.~(\ref{incompressibility condition of x}), the matrix ${\bf S }$ is a singular matrix, i.e, the rank of ${\bf S}$ is $n-1$.
This means that the segment densities of $n$ subchains are not independent.  Then, we take the segment densities of the subchains $i \le n-1$ as the independent variables and the segment density of the $n$-th subchain as a dependent one.
In this case, we should calculate the inverse matrix of the $(n-1) \times (n-1)$ block in the upper left part of ${\bf S }$.  Then, we define an $n \times n$ matrix $\widetilde{{\bf S}}^{-1}$, where the $(n-1) \times (n-1)$ block in the upper left part is filled with the inverse matrix obtained above and the $n$-th row and column are filled with 0 elements.

By using the  $\widetilde{{\bf S}}^{-1}$, the {\bf u} is given by
\begin{equation}
 {\bf u } = - \frac{1}{\beta}\widetilde{{\bf S}}^{-1} {\bf x }.
 \label{equation of u}
\end{equation}

The $\widetilde{{\bf S}}^{-1}$ in eq (\ref{equation of u}) is the exact and general expression of the inverse scattering function derived with the RPA. It should be noted that the matrix $\widetilde{{\bf S}}^{-1}$ is time-independent for the canonical ensemble system.
Therefore, in practical calculation, there is no need to update the $\widetilde{{\bf S}}^{-1}$ for each time step of the dynamical simulations.

\subsection{Subchain scattering function matrix}
\label{Subchain scattering function matrix}

We denote the indices of the subchains in the $p$-type polymer by $i'$ and $j'$. Then the scattering function from the $p$-type polymer which is assumed to be an ideal Gaussian chain is given by
\begin{eqnarray}
\mathcal{S}_{i'i'}^{0p} ({\bf q}) 
&=& \frac{1}{N^p} \int_0^{N_{i'}^{(p)}} ds \int_0^{N_{i'}^{(p)}} ds' \exp(-\frac{ {\bf q} ^ 2 }{6} |s-s'| b^2 ) = \frac{2 N_{i'}^{(p)}}{N^{(p)} x^2} (e^{-x} - 1 + x ) \\
\mathcal{S}_{i'j'}^{0p} ({\bf q}) 
&=& \frac{1}{N^{(p)}} \int_0^{N_{i'}^{(p)}} ds \int_0^{N_{j'}^{(p)}} ds' \exp(-\frac{ {\bf q} ^ 2 }{6} |s-s'| b^2 )
\nonumber \\
&=& \frac{N_{i'}^{(p)} N_{j'}^{(p)} e^{-z}}{N^{(p)} x y} (e^{-x} - 1)(e^{-y} - 1)\ \ \ \ \ \ \ \ (i' \ne j'),
\end{eqnarray}
where $x$, $y$, and $z$ are given by
\begin{equation}
x \equiv R_{Gi'}^2 |{\bf q}|^2,\ y \equiv R_{Gj'}^2 |{\bf q}|^2,\ {\rm and\ } z \equiv R_{Gi'j'}^2 |{\bf q}|^2.
\end{equation}
$R_{Gi'}$ and $R_{Gj'}$ are the radii of gyration ($R_G=(1/6)Nb^2$) of the $i'$-th and $j'$-th subchains, respectively, and $R_{Gi'j'}$ is the averaged radius of gyration of subchains which connect the $i'$-th and $j'$-th subchains. 

As a result, the explicit expression of the $S_{ij}^0 ( {\bf q} )$ is obtained as a product of the average volume fraction of the $p$-type polymer in the system $\bar{\phi_p}$ and the $\mathcal{S}_{i'j'}^{0p}({\bf q})$ as
\begin{eqnarray}
S_{ij}^0({\bf q}) &=& \bar{ \phi_p } \mathcal{S}_{i'j'}^{0p}({\bf q})\\
\label{S0ij}
i&=&\sum_{p'}^{p-1} n^{(p')} + i',
\label{eq of i}
\end{eqnarray}
where ${\bf S }^0$ is a symmetric matrix that satisfies $S_{ij}^0 ( {\bf q} )=S_{ji}^0 ( {\bf q} )$ and the diagonal element $S_{ii}^0 ( {\bf q} )$ is the self scattering function of the $i$-th subchain. The subscript index $i$ for subchains in the system is expressed by using the subscript index $i'$ for subchains in the polymer as eq~(\ref{eq of i}).

\section{Variable time mesh technique for the dynamic density functional theory}
\label{Variable time mesh technique for the dynamic density functional theory}

We introduce a dynamic density functional technique, which is basically a technique for the minimization of the free energy. 
Therefore, when we use the dynamic simulation for the purpose of searching for the local equilibrium structure, a faster convergence is desirable because we do not concern the actual time evolution of the system.
For this purpose, we propose an algorithm for the acceleration of the dynamic density functional theory as follows.

We set the initial time step width $\Delta t$ for a simulation and then introduce a time step ratio $r_t$ ($r_t>1$) and the upper limit of the time step width $\Delta t_{max}$ which is now treated as a variable.
Then, we modify the time step width $\Delta t$ according to the following algorithm.
\begin{enumerate}
  \item[(1)] Set the initial $\Delta t$.

  \item[(2)] Perform one step calculation of the dynamics simulation.

  \item[(3)] If some of the segment densities become negative, then return to the state just before the previous step (2) is performed.  Then, change $\Delta t$ as $\Delta t \leftarrow \Delta t / r_t$ and recalculate step (2). Otherwise, go to step (4).

  \item[(4)] Change $\Delta t$ as $\Delta t \leftarrow \Delta t \cdot r_t$.  If $\Delta t$ exceeds $\Delta t_{max}$ then $\Delta t \leftarrow \Delta t_{max}$.

  \item[(5)] Increase the time of the simulation system by $\Delta t$ and go to step (2).
\end{enumerate}
This acceleration technique becomes effective for a structural change of a system in its early stage and also in the very late stage after the system almost reaches the equilibrium structure. A large $\Delta t$ is automatically chosen for such early and very late stages.

For the purpose of a rough estimation of the equilibrium structure, this acceleration technique combined with the dynamic GRPA theory is useful and efficient compared to the full dynamic SCF theory because the SCF theory demands many iterations for a large $\Delta t$ value.

\section{Efficiency of the hybrid theory and parallel computations}
\label{Efficiency of the hybrid theory}

The hybrid theory improves the efficiency of the SCF calculations by several times as shown in Fig. 6. This acceleration factor is comparable to that of a parallel computation. On the other hand, the hybrid theory requires more memory space than the SCF theory because it needs both the GRPA and SCF calculation schemes. However, the memory space needed for the GRPA theory is roughly proportional to the product of the system size and the number of the subchains.  This required memory space is negligible compared to the memory space needed for the SCF theory, which is in proportion to the product of the system size and the sum of the subchain lengths. Thus the ratio of the memory space GRPA/SCF is about (number of the subchains)/(sum of the subchain lengths).
Thus the hybrid theory can achieve a few times faster simulation with almost the same size of the required memory as the SCF theory.  This is the advantage of the hybrid theory compared to the parallel computation of the SCF theory, which requires a large memory space for parallelization.

\newpage

\bibliographystyle{tipsj}

\begin{figure}[H] \begin{center} \includegraphics[width=0.8\linewidth]{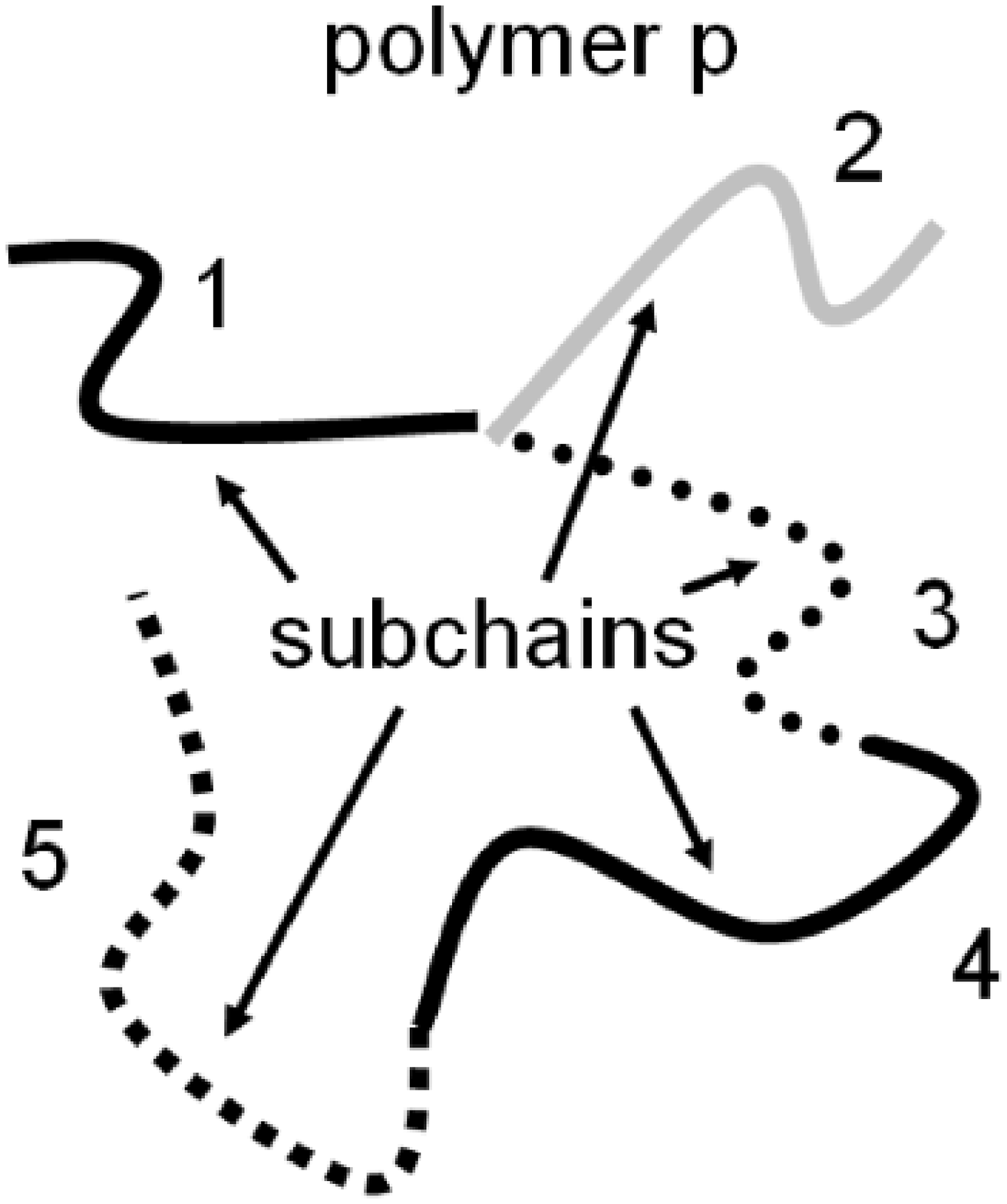} 
\caption{ } \label{Polymer} \end{center} \end{figure}

\begin{figure}[H] \begin{center} \includegraphics[width=0.8\linewidth]{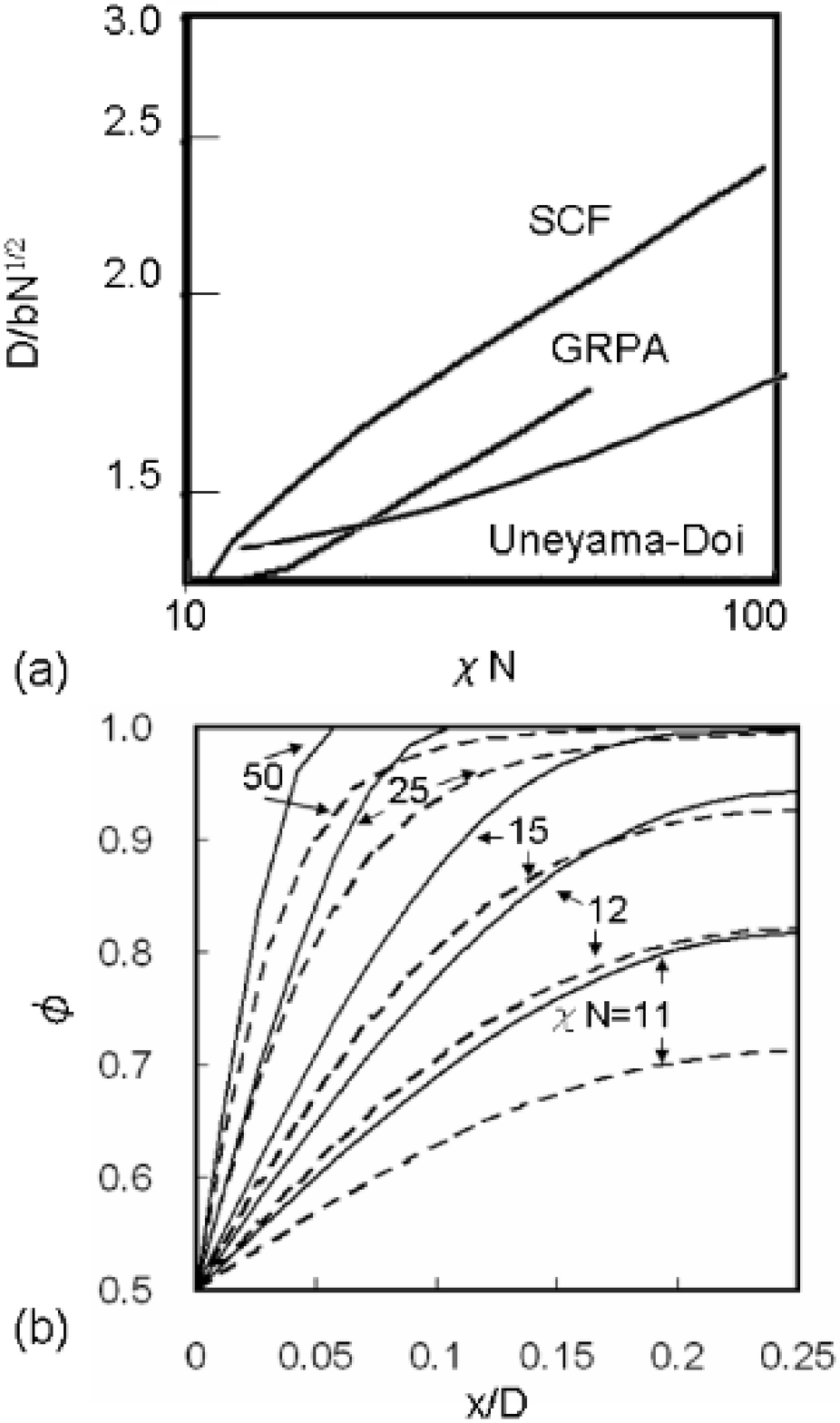} 
\caption{ } \label{GRPA} \end{center} \end{figure}

\begin{figure}[H] \begin{center} \includegraphics[width=1.0\linewidth]{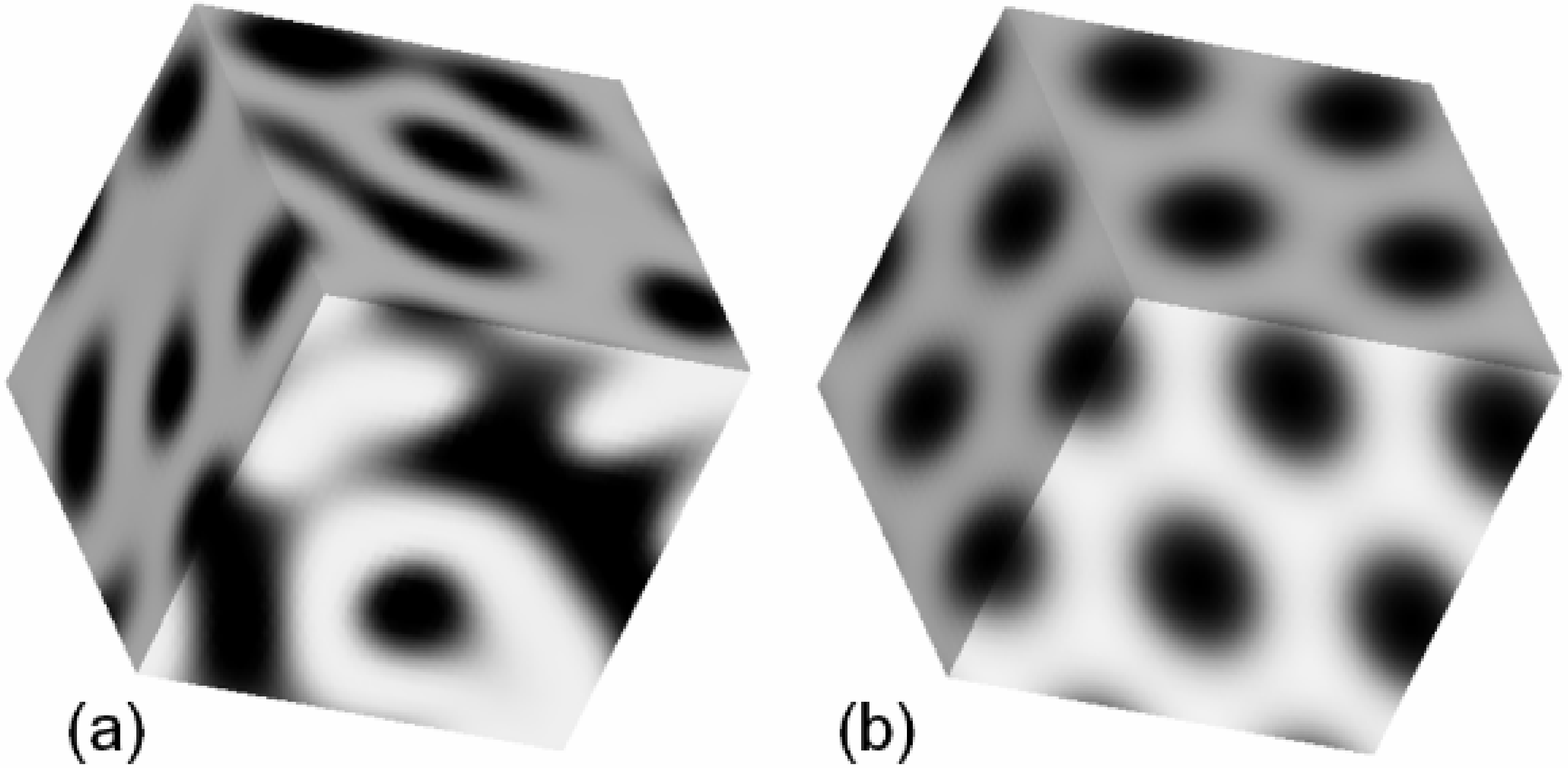} 
\caption{ } \label{Cylinder} \end{center} \end{figure}

\begin{figure}[H] \begin{center} \includegraphics[width=1.\linewidth]{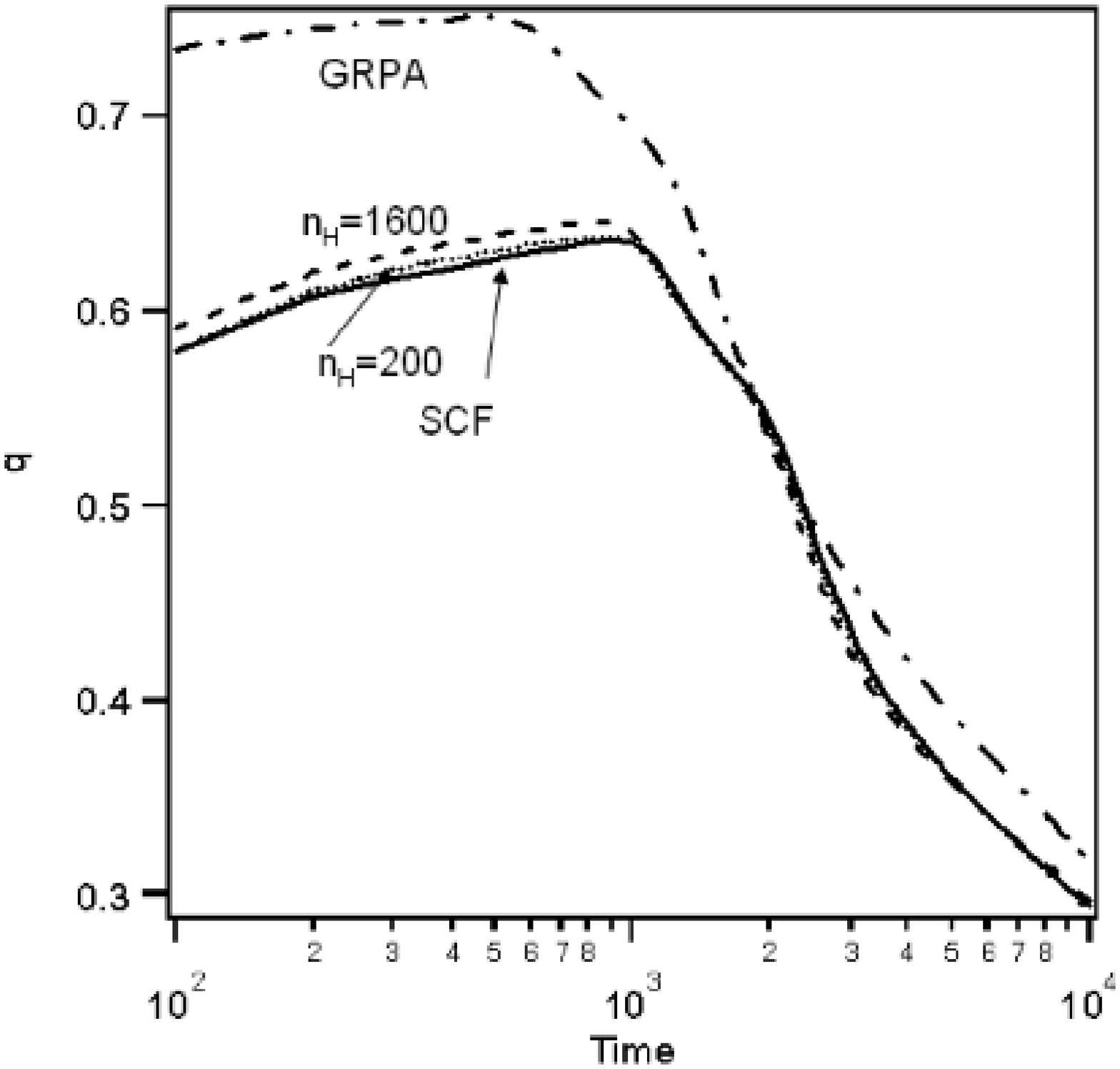} 
\caption{ } \label{Blend} \end{center} \end{figure}

\begin{figure}[H] \begin{center} \includegraphics[width=1.0\linewidth]{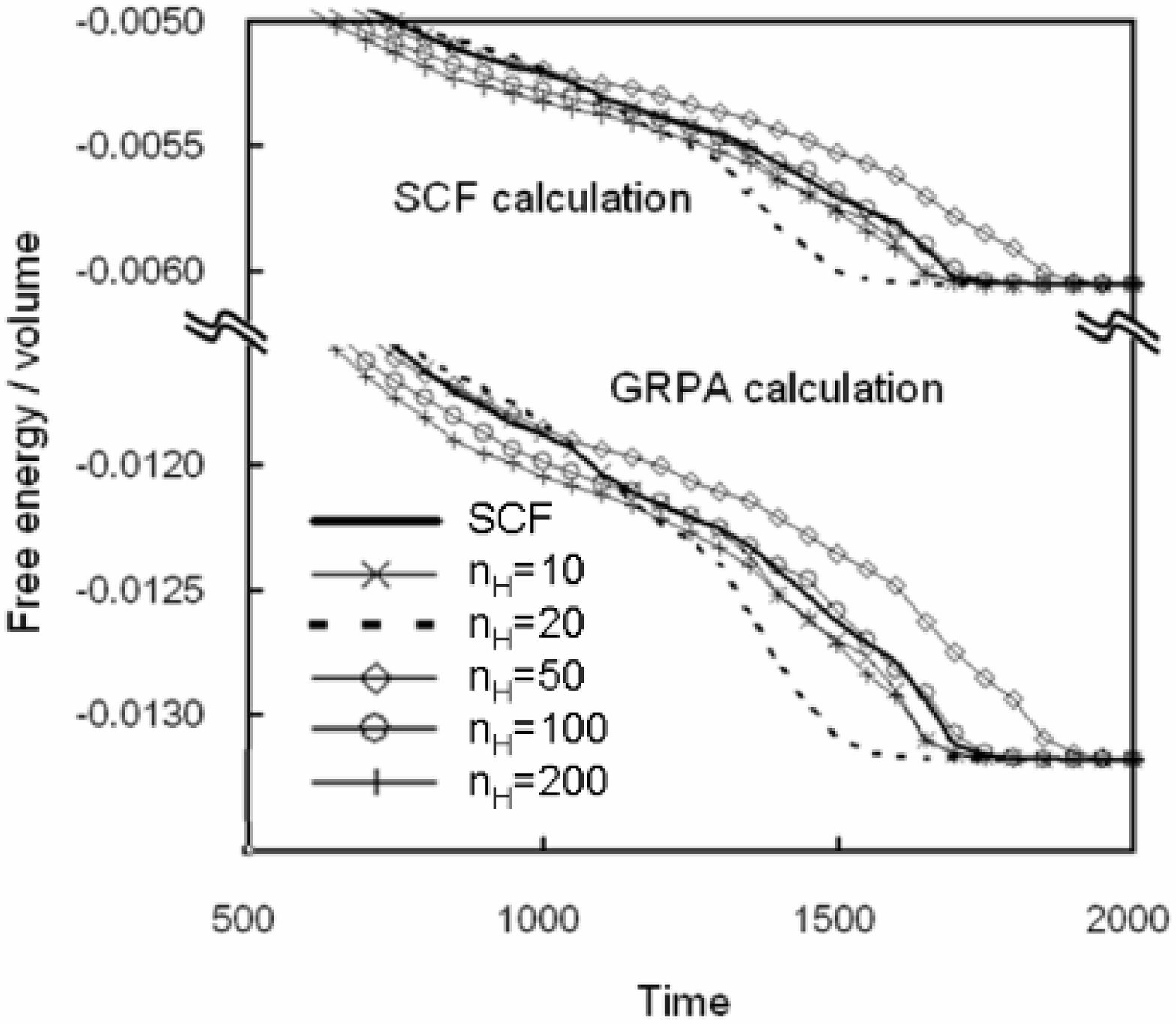}
\caption{ } \label{Cylinder_FE} \end{center} \end{figure}

\begin{figure}[H] \begin{center} \includegraphics[width=1.0\linewidth]{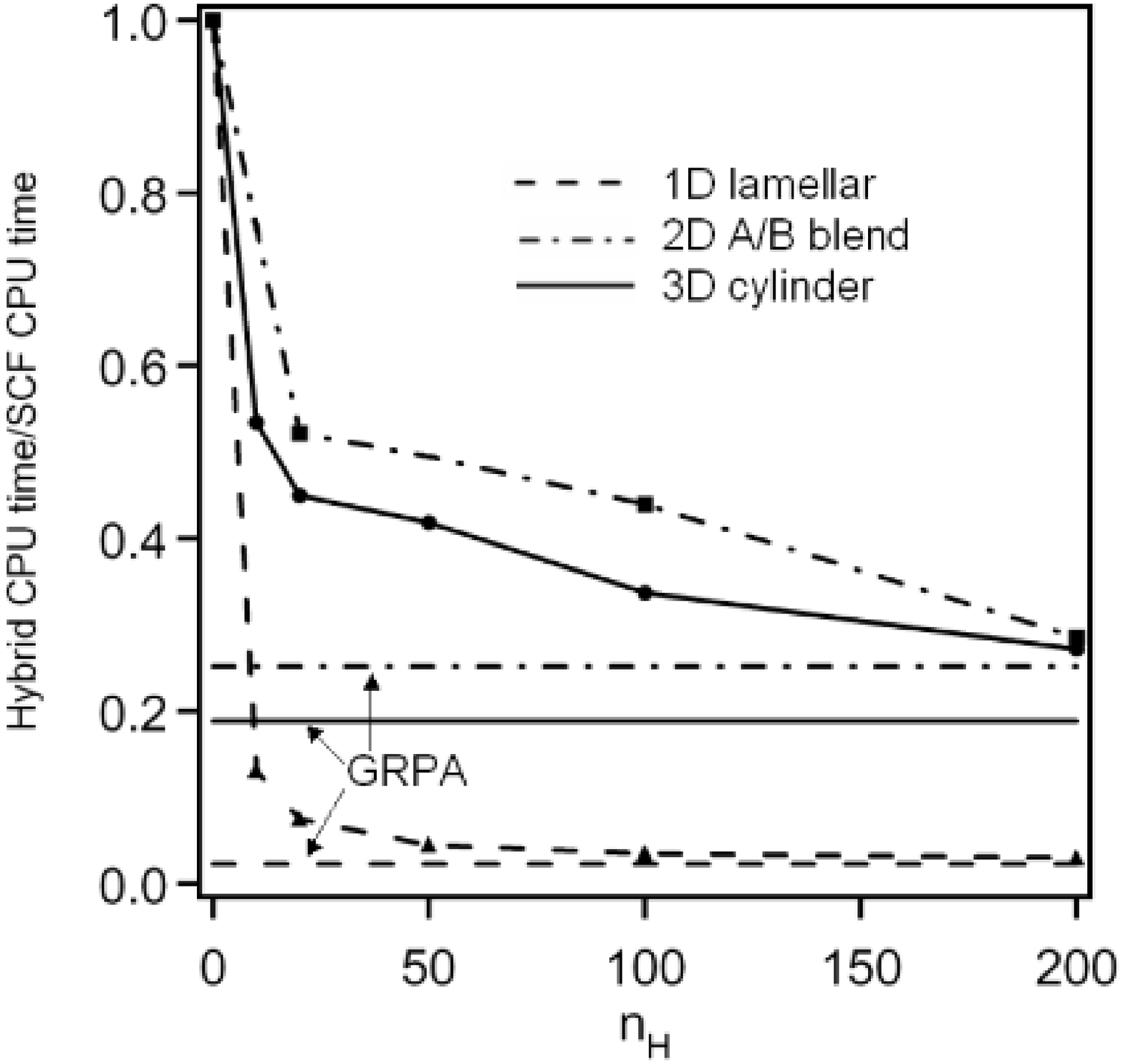} 
\caption{ } \label{computational_time} \end{center} \end{figure}

\begin{figure}[H] \begin{center} \includegraphics[width=0.8\linewidth]{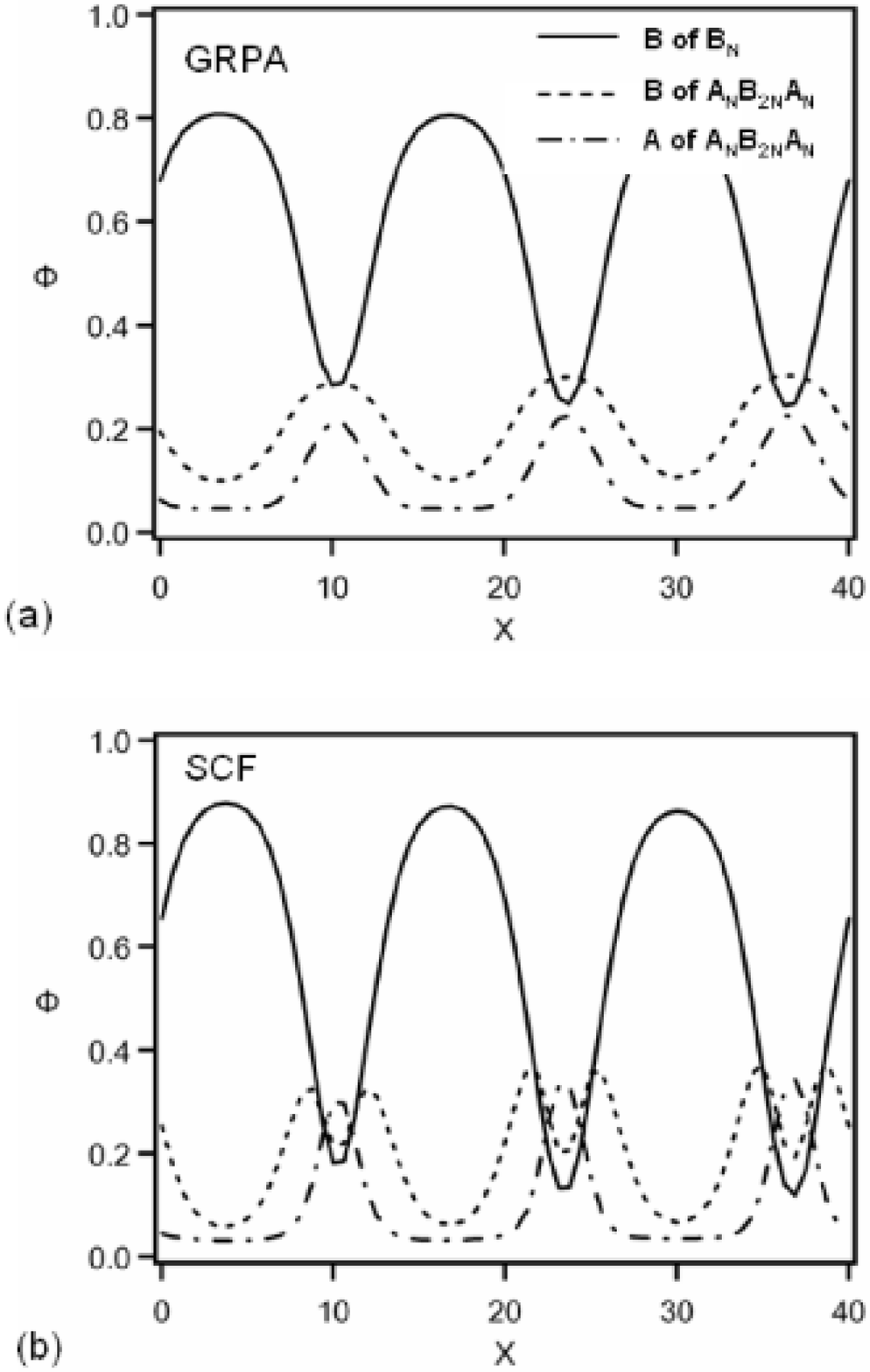} 
\caption{ } \label{Exception} \end{center} \end{figure}

\begin{figure}[H] \begin{center} \includegraphics[width=0.8\linewidth]{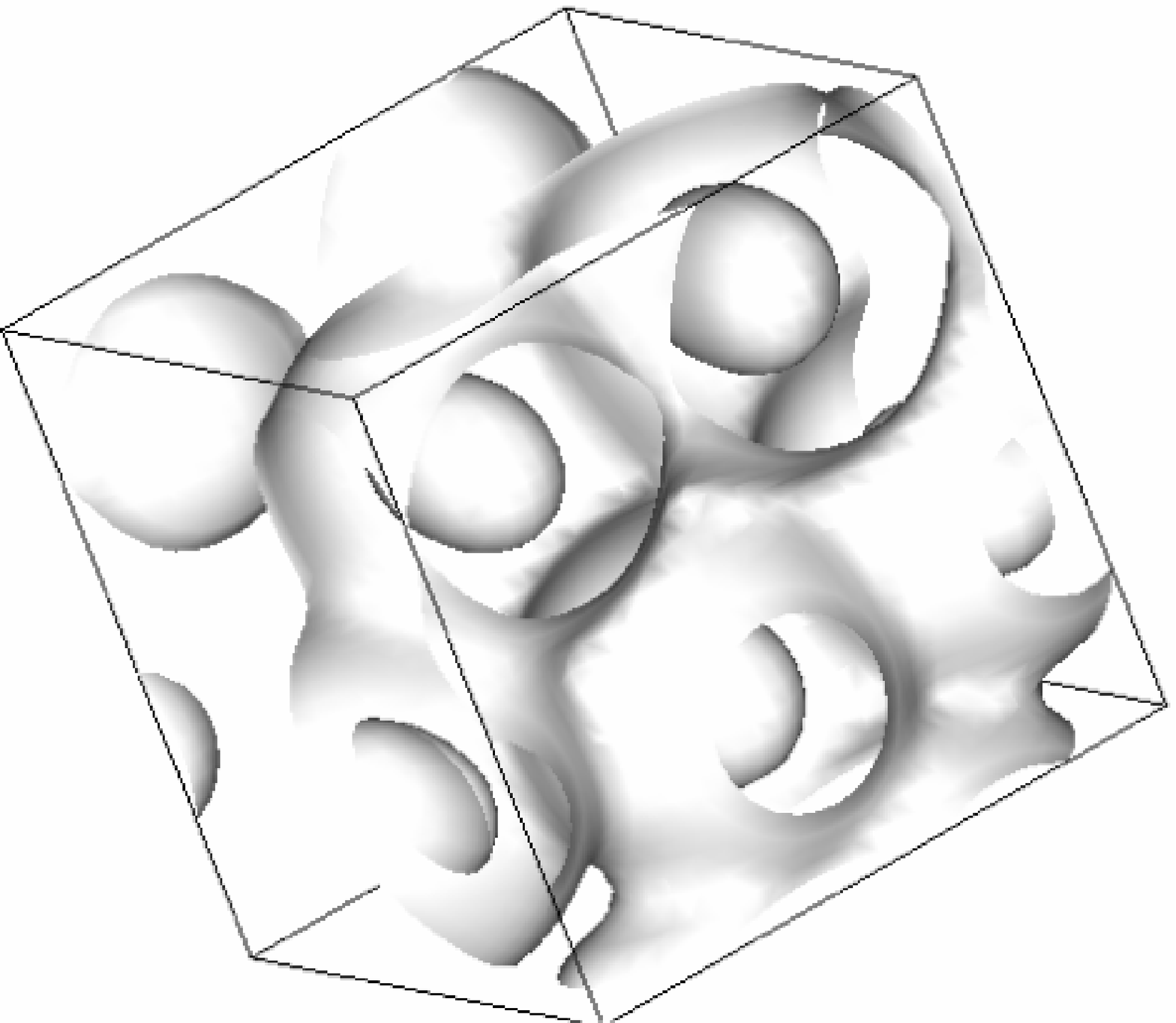} 
\caption{ } \label{B_ABA2000} \end{center} \end{figure}

\begin{figure}[H] \begin{center} \includegraphics[width=1.0\linewidth]{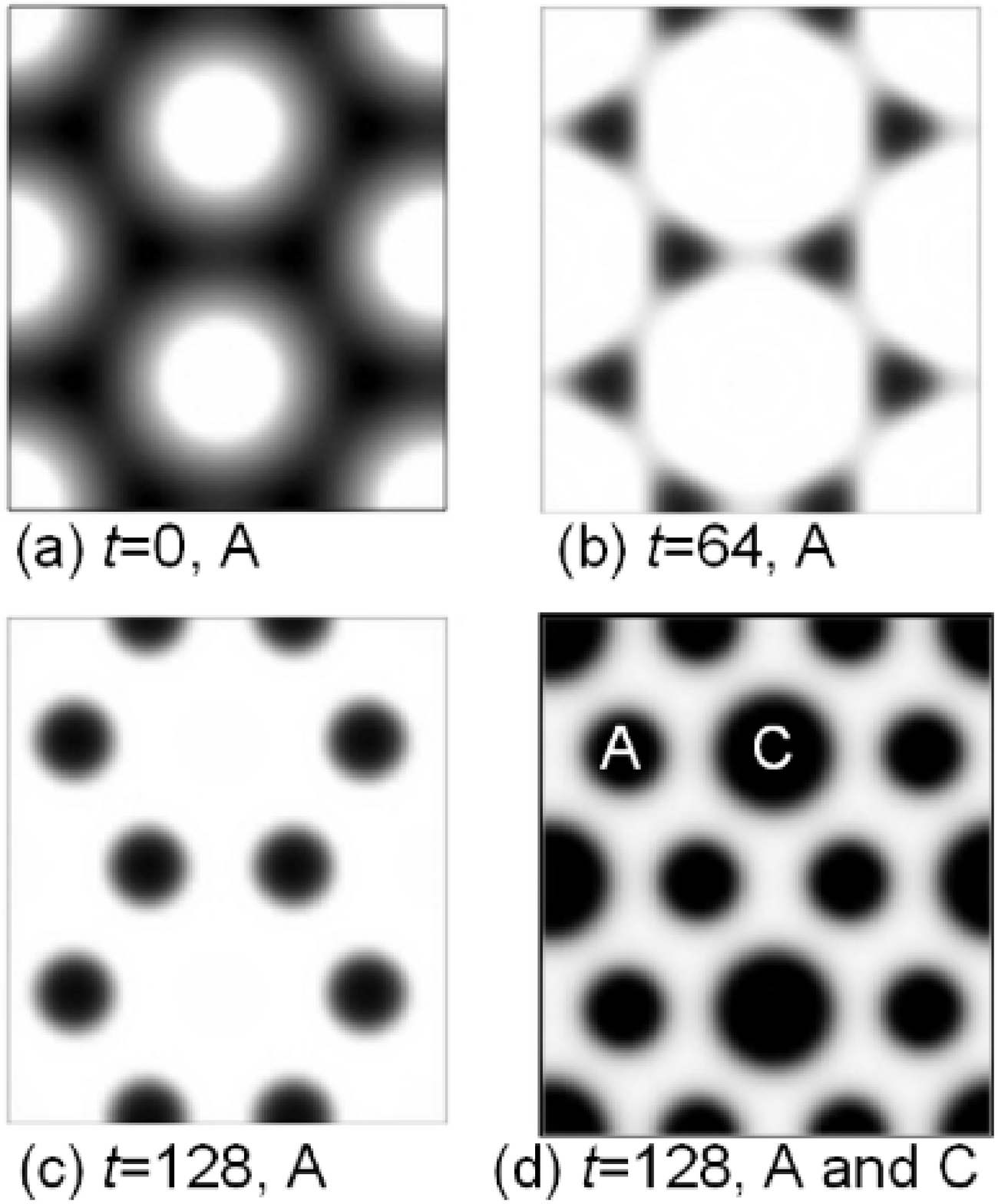} 
\caption{ } \label{ABC} \end{center} \end{figure}

\newpage
{\Large\bf Figure captions}
\newline\newline
Figure \ref{Polymer}.
A schematic picture of a polymer architecture treated in this article.
\newline\newline 
Figure \ref{GRPA}.
Results of an one-dimensional GRPA calculation on a symmetric AB diblock copolymer.
(a) The simulation results of the SCF, GRPA, and Uneyam and Doi's theories
\cite{Uneyama Doi}
on the equilibrium periods $D$ of the lamellar structure are shown for various values of the $\chi N$.
(b) A comparison between the segment density profiles of the GRPA theory (solid lines) and those of the SCF theory (dashed lines) for the lamellar structure is shown.
\newline\newline
Figure \ref{Cylinder}.
A comparison of the domain structures of a melt of an AB diblock copolymer with $N=20$, $f=0.35$, $\chi N=15$ calculated with the dynamic GRPA and the hybrid theories in the late stage is given.
The spatial mesh size is $32^3$ with the mesh width $\Delta x$ = 0.5375.
The values of $f$ and $\chi N$ correspond to those inr the stable region of the hexagonally packed cylinder (HEX) structure on the phase diagram of the SCF calculation
\cite{Matsen Schick}.
\newline\newline 
Figure \ref{Blend}.
The time dependence of the wave number at the peak position of the scattering functions of an A/B polymer blend that undergoes spinodal decomposition is given.
The system is a symmetric polymer blend with $N_A=N_B=10, \bar{\phi_A}=\bar{\phi_B} = 0.5$ and $\chi N=3$.
The spatial mesh size is $256^2$ with the spatial mesh width $\Delta x$ = 1.0 and the time evolution is integrated with the time mesh width $\Delta t = 0.01$.
\newline\newline 
Figure \ref{Cylinder_FE}.
Time dependence of the free energy densities for the micro phase separation of an AB diblock copolymer melt is shown.
The simulation condition is the same as that of the system shown in Fig. \ref{Cylinder}.
Shown are the results of the hybrid calculations with several values of $n_H$ with $\Delta t = 0.01$.
The free energy density is evaluated with both the SCF (upper set of curves) and GRPA (lower set of curves) theories to show that these two theories give qualitatively the same results.
\newline\newline 
Figure \ref{computational_time}.
Dependence of the performance of the hybrid theory on the model parameters and the used techniques is shown.
The computational times for individual runs are normalized using that for the full SCF calculation ($n_H=1$).
Results for three types of simulations with $\Delta t = 0.01$ are indicated:
(a) a one dimensional diblock copolymer system with $N=100$, $f=0.5$, $\chi N=15$, and the number of grid points = 64 with mesh width $\Delta x$ =  0.25,
(b) a two dimensional A/B polymer blend system whose simulation condition is the same as that in Fig. \ref{Blend},
and 
(c) a three dimensional diblock copolymer system whose simulation condition is the same as that in Fig. \ref{Cylinder}.
The horizontal lines indicate the necessary computational time for the dynamic GRPA theory, which gives the lower limit of the hybrid simulations.
\newline\newline 
Figure \ref{Exception}.
A comparison of the segment density profiles between the dynamic GRPA and dynamic SCF theories for an exceptional case that requires small $n_H$ for the hybrid theory is given.
The system is a one dimensional polymer blend composed of a homopolymer $B_N$ and a triblock copolymer $A_NB_{2N}A_N$ with $N=10$ and $\bar{\phi}_B = 0.6$. (a) The segment density profile obtained with the GRPA theory and (b) that obtained with the SCF theory are shown. In the latter case, a fine structure inside the interfacial region can be observed, which cannot been reproduced in the former case.
\newline\newline 
Figure \ref{B_ABA2000}.
A domain structure of a three dimensional $B_N/A_NB_{2N}A_N$ blend obtained with the hybrid theory with $n_H=5$ at $t=2000$ is shown. The condition of the system is the same as that of Fig. \ref{Exception}.  The simulation is done on a 3-dimensional mesh with $40^3$ mesh points. The isosurfaces of the B segment density of the ABA copolymer at the value 0.2 are shown.
\newline\newline 
Figure \ref{ABC}.
Time dependence of the formation of the domain structure of a melt of an $A_{20}B_{60}C_{20}$ polymer starting from an equilibrium HEX structure of the C segments is shown. (a) The initial distribution of A segments (black) where C segments form the HEX structure and, A and B segments form the matrix phase. The used parameters are $\chi_{AB} = 0.$, and $\chi_{AC} = \chi_{BC} = 0.4$, and the system size is $30.5 \times 35.2$, which is divided into $64^2$ meshes.  In order to adjust the system size to the natural period of the HEX structure, we used the SSO method
\cite{G to C}.
At $t=0$ the $\chi$-parameters are suddenly changed to $\chi_{AB} = \chi_{AC} = \chi_{BC} = 0.4$.  Then, the phase separation process is simulated with $n_H$=5 (Figs.~(b) and (c)).  
The A segment distributions at (b) $t=64$ and (c) $t=128$ are shown, respectively.
Figure~(d) is the same system as in Fig.~(c) but both the A and the C segments are shown simultaneously.  In this domain structure, the A segments form a spatial motif around the HEX structure of the C segment.

\newpage
\newpage
\newpage
\newpage
\newpage
\newpage
\newpage
\newpage
\newpage

\end{document}